\documentclass[11pt]{article}

\usepackage[]{graphics,graphicx}

\usepackage{pstricks}
\usepackage{amsthm}
\usepackage{amsmath}
\usepackage{blindtext}

\usepackage{multicol}
\usepackage{parskip}

\usepackage{graphicx}
\usepackage{graphics}

\usepackage{float}
\usepackage{setspace} 

\RequirePackage{lineno}

\begin{document} 



\begin{figure*}[t!]
\begin{center}
\hspace{5mm}
 \includegraphics[width=9cm]{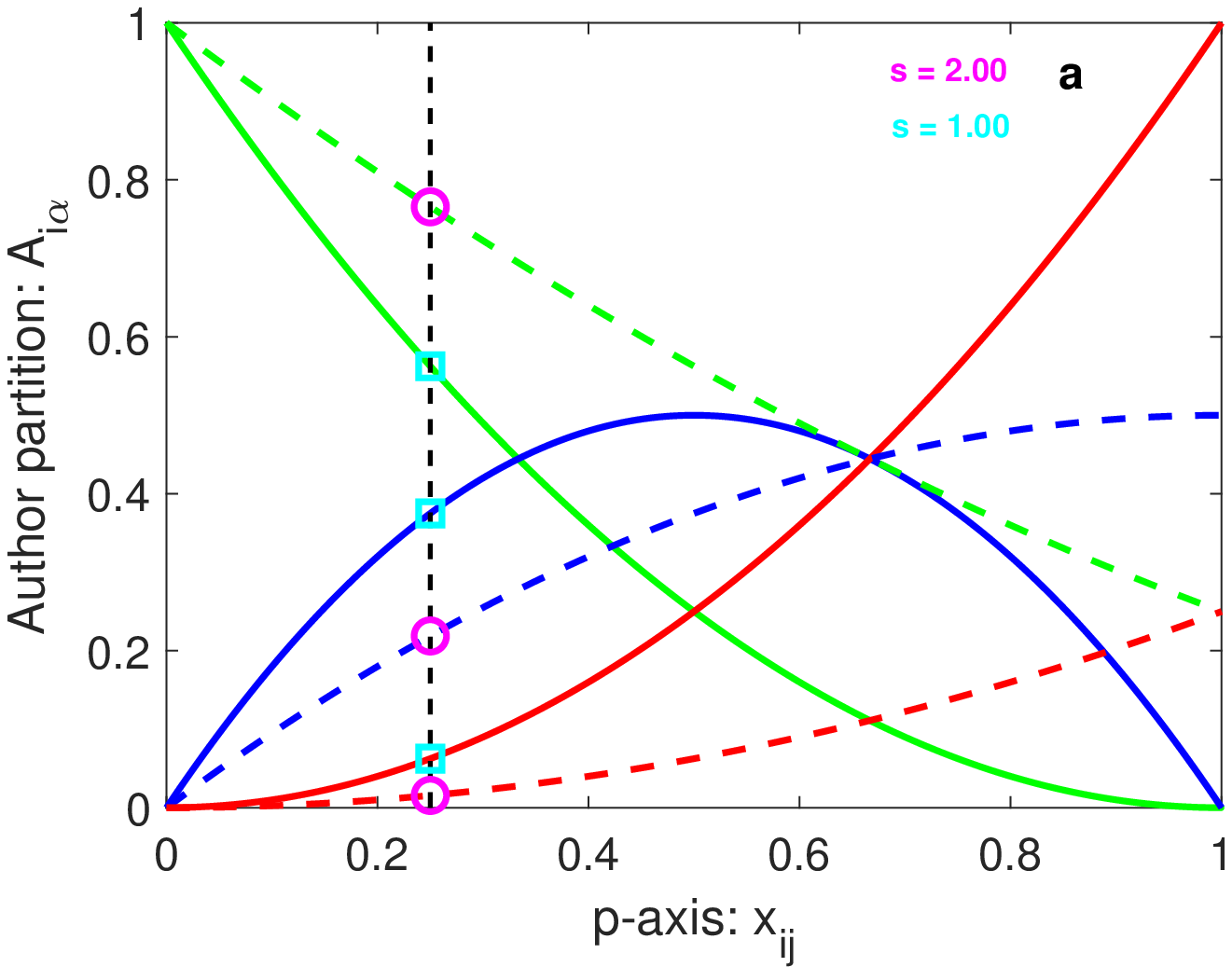}
 \hspace{-5mm}
 \includegraphics[width=9cm]{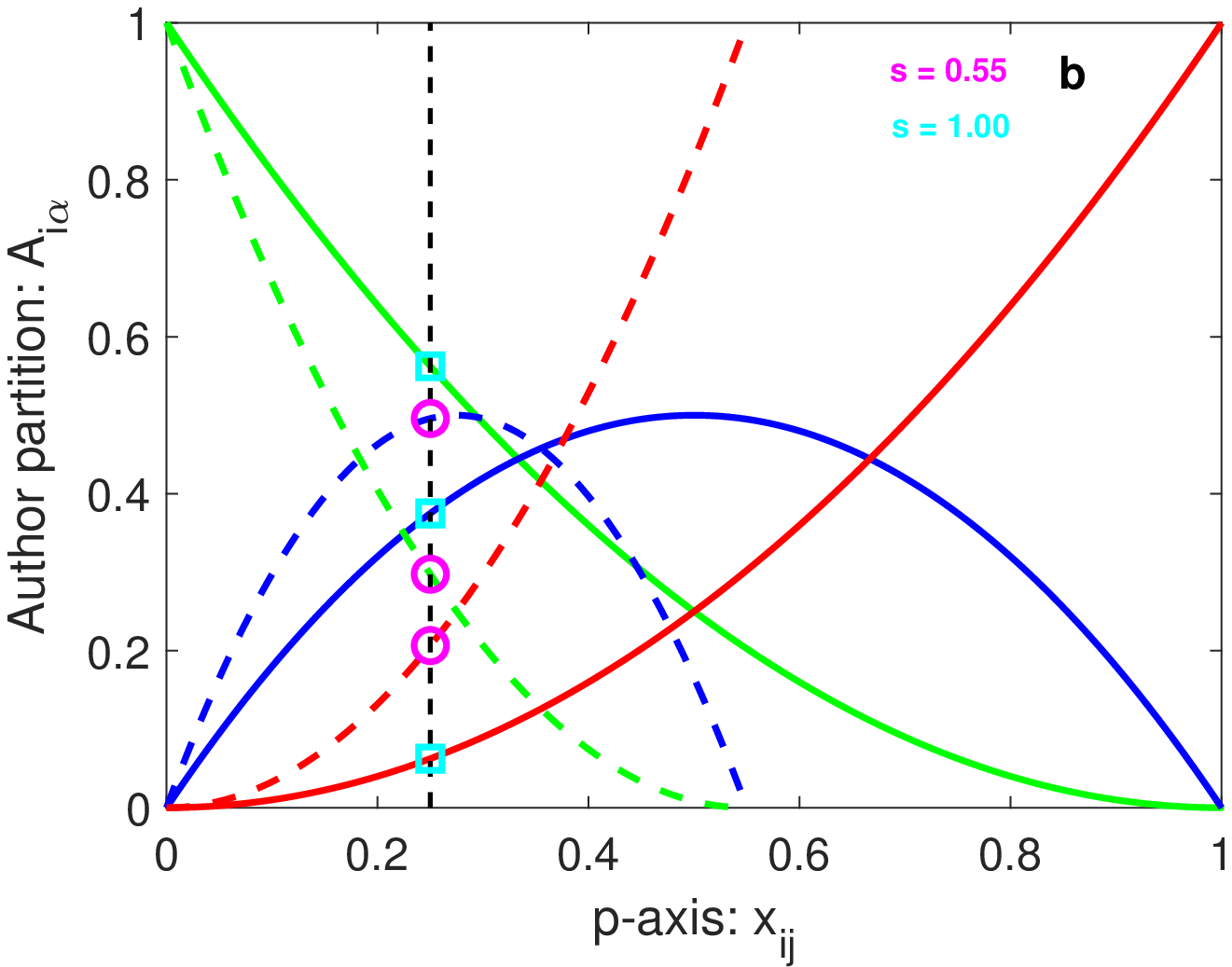}\vspace{-3mm}
  \end{center}
  \caption{(a) The stretched Bernstein-S polynomials (dashed lines), (b) contracted Bernstein-S polynomials (dashed lines) against the Bernstein-S polynomials (solid lines). Partitioning series generated by the p-axis (vertical dashed line in gray) as it intersects Bernstein polynomials (square symbols in cyan) and the Bernstein-S polynomials (circle symbols in magenta), producing fundamentally different author contribution partitionings.}
  \label{Fig_s}
\end{figure*}

\begin{center}
{\LARGE\bf P-Index}
\\[10mm]
{\large Shiva P. Pudasaini
\\[3mm]
Technical University of Munich\\
School of Engineering and Design, Chair of Landslide Research}\\
{Arcisstrasse 21, D-80333, Munich, Germany}\\[1mm]
{E-mail: shiva.pudasaini@tum.de}\\[7mm]
\end{center}
\noindent
{\bf Abstract:\\
I propose the P-Index that genuinely constitutes a well-defined, compact author citation metric.}
\begin{multicols}{2}
{\it {\bf \large Significance:}  By considering the actual author contributions, the P-Index comprehensively resolves the question of true author citation measure. The P-Index brings a complete overhaul in our perception of the author citation metric, including total citations, from which the researchers and authors greatly benefit with compact measures. It is anticipated that the P-Index leads to a fuller appreciation of who made how much real contributions by providing the unique author citation metric. This principle is expected to roll out swiftly across the scientific and academic disciplines as the P-Index can equally be applied to any institutions, including universities, research institutions, academic fields, and nations.}
\\[3mm]
Authorship grants merit for a researcher's contribution to a study and carries accountability for published work, and has important scientific, academic, social, and financial inferences to the author (ICMJE, 2018). As the collaborative and team-based research is increasing, the number of authors per article is rising across scientific fields (Lapidow and Scudder, 2019). Many, but not all, journals individually outline 
{what it means to be a contributing author and} 
the nature of each author’s role. 
\\[3mm]
 The number of co-authors in a scientific research article may vary from one to several thousands, and currently, tens to hundreds of co-authors is not that unusual in a hyperauthorship paper (King, 2012; Castelvecchi, 2015; Lapidow and Scudder, 2019), and such papers are blooming. The order of author names on a publication is commonly used in most scientific communities to award them recognition and designate responsibility. 
 With exceptions that authors are presented in alphabetical order, it is assumed that the first author contributes the most. Intellectually and financially, the last author can be propelling the research (Lapidow and Scudder, 2019). 
\\[3mm]
However, authors contributions cannot just be determined by their positions in the byline. To-date, no firm guidelines exist to ensure an honest and unbiased explanation of authors’ contributions (Abambres and Arab, 2016), and at most, practically only qualitative descriptions are in use. Authors are expected to meet some norms. The International Committee of Medical Journal Editors (ICMJE, 2018) outlines four criteria, that all must be substantially fulfilled by an individual to be a co-author of an article: Conception or design of the work, or the acquisition, analysis, or interpretation of data; drafting the work or revising it; final approval of the version; agreement to be accountable for all aspects of the work. The Proceedings of the National Academy of Sciences (PNAS) and Nature journals adopt the authorship criteria from McNutt et al. (2018). Similarly, in Science and other journals, often the author contributions include conceptualization, methodology, investigation, visualization, funding, supervision, and writing.   Some journals (e.g., Lancet) mention about equal contributions of some co-authors and some shared first authorship (Lapidow and Scudder, 2019). However, none of the authors, journals and publishers have yet quantified how much a particular author contributed to an article, making it impossible to calculate that person's real total citations, the citation index, and the author metric in a database.  
\\[3mm]
To quantify an individual's scientific research output, Hirsch (2005) proposed the H-Index. It measures one's research productivity and impact, even more so the performance of a journal.
 This informs that a particular author (or a journal, or institution) has at most H number of articles with at least H number of citations, which is analogous to the Eddington Number (cycling or walking). However, the H-Index does not tell us about the actual number of total citations, and the citation index of an individual, reflecting that person's genuine personal contributions to all the articles this individual has co-authored. 
Often, the H-Index may not correctly mirror the strength of a researcher, scientist or an author, similarly academia, research institution or a journal, because it does not have enough intrinsic controls of genuine contributions of an author or institution, and may even be exploited or influenced for desired outcomes.  
Here, I propose a novel measure, called the P-Index, the Personal-Index, that uniquely and truly explains a researcher's total citations, {and the citation index} in a database. This provides us with a completely new perspective and {an unified} measure of the author citation {metric that} one really deserves. 
\\[5mm]
{\bf \large The P-Index}
\\[0mm]
The existing scientific research indices do not take into account one's actual scientific contributions. It might not be universally true, but, usually, the authors are arranged from the first place to the second, and so on, according to their contributions. Often, the first author might have contributed a lot as compared to other co-authors. These aspects must be considered properly with clear measurable values, which, however, have never been done
{in a unified, compact and automated manner}, although there have been some efforts (Verhagen et al., 2003; Tscharntke et al., 2007). The best way to do this is numerically partitioning the total citations of the paper into all the co-authors according to their actual contributions {(Galam, 2011)}. Since the paper with a large number of authors is expected to garner more citations, an appropriate partition of citation of each paper over all authors is required. As the author metrics are becoming more and more important in scientific and academic communities, here, I formally propose the P-Index that provides an author citation metric, a unique, {unified and compact} measure solely depending on author's personal contributions on all publications. It is based on the simple principle of harvesting in proportion to one's personal contributions and outcomes.
\\[3mm]
I aim to construct a well-defined and well-justified, automated author citation metric. Let an author, or a research person, $A$ has published $N$ number of articles (articles, papers, or citable literature) listed in a database, with $N_s$ number of papers as a single author, and $N-N_s$ number of papers with multiple authors. Let $i$ denote the paper number (may be arranged in chronological order), $C_i$ is the total number of citations of the $i$th paper coauthored by $M_{i_j}$ number of authors ($j$ indicates the number of additional coauthors for $i$th paper). For a paper with multiple authors, usually each individual contributes differently. Let, $A$'s contribution to the paper $i$ be ${{i}_{\alpha, j}}$  (with $\alpha = 0, ..., j$), a unique numerical value which lies in $(0, M_{i_j}]$. For very low contribution ${i}_{\alpha, j} \to 0$, and for very high contribution ${i}_{\alpha, j} \to M_{i_j}$. For the single author paper, ${i}_{\alpha, j} = M_{i_j}$. Then, based on $A$'s actual contributions over all the papers, total citations $C$ for $A$ can be rigorously defined as:
\begin{eqnarray}
{\small
 C =  \sum_{i =1}^{N_s} C_i + \!\!\!\sum_{i = N_s+1}^{N} \!\frac{{i}_{\alpha, j}}{M_{i_j}} C_i
= C_{N_s}^s +\!\!\! \sum_{i = N_s+1}^{N} \!\frac{{i}_{\alpha, j}}{M_{i_j}} C_i,
\label{Eqn_1}
}
\end{eqnarray}
where, $C_{N_s}^s$ is the total citations with the single-authored articles for $A$. Note that, for each article $i$, $M_{i_j}$ is fixed. However, $i_{\alpha, j}$ varies over all the authors for that particular paper $i$. I define the author contraction $\mathcal A_{i_{\alpha, j}}: = {i}_{\alpha, j}/{M_{i_j}}$, as the actual author contribution factor for $A$, a precise numerical value, for the multiauthor article $i$. Then, (\ref{Eqn_1}) can be rewritten as:
\begin{equation}
C = C_{N_s}^s + \sum_{i = N_s+1}^{N} \mathcal A_{i_{\alpha, j}} C_i.
\label{Eqn_2}
\end{equation}
{There are two important aspects we must consider in generating the actual author citation metric. First, it must provide a justifiable representative value. Second, for an author, there can be several articles without any citations, or with negligible citations. Both of these aspects can be assured by taking the average over all the articles in (\ref{Eqn_2}) with the number of articles $N$. This results in}
\begin{equation}
Q
= \frac{1}{N}\left[ C_{N_s}^s + \sum_{i = N_s+1}^{N} \mathcal A_{i_{\alpha, j}} C_i\right ].
\label{Eqn_2PQ}
\end{equation}
{For someone with very few articles, most probably being the first or around the first author, each article with high citations, the $Q$ value is relatively high, although that person's productivity is low (low production, high personal impact). In contrast, another person might have numerous articles, for some being the first or around the first authors, but for many articles somewhere in between, or closer to the end in the byline (high production, low personal impact). This person is highly productive, but may get a lower $Q$ value than the first person. So, we need to balance between the productivity and real personal impact. This is accomplished by taking the measure:
\begin{equation}
P
= \min\{ N, Q\}.
\label{Eqn_2P}
\end{equation}
This rationally assigns the higher $P$ value to the individual who has a high number of impactful articles with higher personal contributions. For most authors, $P$ and $Q$ can be the same.}
 I call this P the Personal-Index, or the P-Index, representing the actual {author index, based on the} total number of citations over all the articles the author $A$ has contributed to the database. {In practice, if we wish, we can round $P$ to its nearest integer value, $P: = \mbox{round}(P)$, which is non-negative. The same principle can be applied to $C$.}
\\[3mm]
Depending on each author's actual contribution to the publication $i$, by definition, the author contraction $\mathcal A_{i_{\alpha, j}}$ varies between 0 and 1, i.e., $\mathcal A_{i_{\alpha, j}} \in (0, 1] $. 
If there are $j+1$ authors for the article $i$, then we must have
\begin{equation}
\sum_{{\alpha} = 0}^{j} \mathcal A_{i_{\alpha, j}} 
= \frac{1}{M_{i_j}}\sum_{\alpha = 0}^{j} {i}_{\alpha, j}
= \frac{1}{M_{i_j}}{M_{i_j}}
= 1. 
\label{Eqn_3}
\end{equation}
Based on their genuine contributions on the paper $i$, the contraction $\mathcal A_{i_{\alpha, j}}$ constitutes the partition of unity over all $j+1$ authors. The existing literature do not provide a person's real contribution measure ${i}_{\alpha, j}$ to the article $i$, but by default set ${i}_{\alpha, j} = {M_{i_j}}$. The existing author citation indexes assign too much contribution to the citation of the paper $i$ by the amount $\left(j +1 \right)C_i$, whereas the actual citation of $i$ is just $C_i$. None of the authors of a multiauthored article can add $C_i$ into their citation metrices, but they can only count $\mathcal A_{i_{\alpha, j}} C_i$.
For the existing literature, one may think to assume a particular situation such that, on average, all authors contributed equally, i.e., ${i}_{\alpha, j}= 1$. 
This superficially may enable us to obtain an easy and quick measure of $A$'s personal total citations and the Personal-Index with $\mathcal A_{i_{\alpha, j}} = 1/{M_{i_j}}$. However, this is inappropriate, because it unjustifiably distributes $C_i$ equally over all authors, irrespective of their actual, and largely differing contributions. 
\\[3mm]
To illustrate the importance of proper partitioning of the total citation of the article $i$ among the authors, assume that article $i$ has 3 authors with respective contributions $\mathcal A_{i_{0, 2}} = 0.65, \mathcal A_{i_{1, 2}} = 0.20$, and $\mathcal A_{i_{2, 2}} = 0.15$. Since, $M_{i_j} = 3$, this can be represented alternatively as $\mathcal A_{i_{0, 2}} = 1.95/3, \mathcal A_{i_{1, 2}} = 0.60/3$, and $\mathcal A_{i_{2, 2}} = 0.45/3$. Moreover, ${i_{0, 2}} + {i_{1, 2}} + {i_{2, 2}} = 1.95 + 0.60 + 0.45 = 3$, and $\mathcal A_{i_{0, 2}} + \mathcal A_{i_{1, 2}} + \mathcal A_{i_{2, 2}} = 1$, which constitutes the partitioning of citations with the partition of unity. This clearly demonstrates that the existing citation measure for the author $A$ (or, $\alpha$) is improper as it sets $\mathcal A_{i_{\alpha, j}}= 1$, resulting in largely excessive measure of citations for all the authors $\alpha$ of $i$, particularly for $\mathcal A_{i_{2, 2}}$ and $\mathcal A_{i_{1, 2}}$. 
However, we need to develop and adopt some mathematically well-defined, well behaving, simple, and automated methods for partitioning author contributions.
\\[4mm]
{\bf \large An automated partition of contributions}
\\[0mm]
Here, I present a method with which the existing deficiency in partitioning author contributions can be removed, and the citation measure can be rectified, even more so for the forthcoming literature. There can be different ways to construct the contraction function $\mathcal A_{i_{\alpha, j}}$. {Galam (2011) suggested partitioning author contributions with some discrete protocols.}
Here, I propose a simple, beautiful, {compact} and automated method that can be applied to any discipline and number of authors with any rule the institutions or the authors define, or prefer. Consider the functions: 
\begin{equation}
\mathcal A_{i_{\alpha, j}} \left ( x_{i_j}\right)= 
\begin{pmatrix}
j\\
\alpha
\end{pmatrix}
x_{i_j}^{\alpha} \left ( 1 - x_{i_j}\right )^{j-\alpha},
\label{Eqn_Partition_1}
\end{equation}
where $\mathcal A_{i_{\alpha, j}}$ are the $j+1$ Bernstein basis polynomials (Bernstein, 1912) of degree $j$, 
{\footnotesize $\begin{pmatrix}
j\\
\alpha
\end{pmatrix}$} are the binomial coefficients, and $x_{i_j} \in [0, 1]$.
For our purpose, $x_{i_j}$ takes a fixed value in $[0, 1]$, which depends on the paper $i$ and the number of co-authors $j$. The Bernstein polynomials in (\ref{Eqn_Partition_1}) constitute a partition of unity for author contributions:
\begin{equation}
\sum_{{\alpha} = 0}^{j} \mathcal A_{i_{\alpha, j}}\left ( x_{i_j}\right)
= 1.
\label{Eqn_Partition_2}
\end{equation}
There are many interesting and important mathematical properties and physical implications of the Bernstein polynomials. These polynomials are symmetric about the vertical line at $x_{i_j} = 0.5$. So, we can restrict our analysis in the half domain $0 \le x_{i_j} \le 0.5$. 
\\[3mm]
The main essence of partitioning lies in selecting $x_{i_j}$. I call it the p-axis, and denote as $p^a_{i_j} = x^a_{i_j}$. With the p-axis, I invented a clear control that can be fixed (recommended, or made mandatory) by any standard (international, national, institutional, or discipline-wise), or decided by the authors themselves. 
This is important. The author contribution partition can change drastically depending on the position of the p-axis. The p-axis is parallel to the ordinate line ($y$-axis). For a single author article, $p^a_{i_0} = 0$, which intersects the trivial Bernstein polynomial, $\mathcal A_{i_{0, 0}}$ = 1 at the ordinate. As the number of authors increases from $1$ to $j+1$, there are $j+1$ number of non-trivial Bernstein polynomials $\mathcal A_{i_{\alpha, j}}$. Then, for any choice of $p^a_{i_j} = x^a_{i_j} >0$, the p-axis $p^a_{i_j}$ vertically intersects the polynomials at $j+1$ number of points, the sum of all the ordinate values {(all positive)} of these coordinate points is unity. We arrange these values in descending order (hereafter, we follow this rule)
$\mathcal A_{i_{\alpha, j}}: = \mbox{descend} \left(\mathcal A_{i_{\alpha, j}}\right)$,
and assign them to the $j+1$ authors, in accordance with their contributions, also arranged in descending order.
\end{multicols}
\begin{figure}[t!]
\begin{center}
\hspace{5mm}
 \includegraphics[width=9cm]{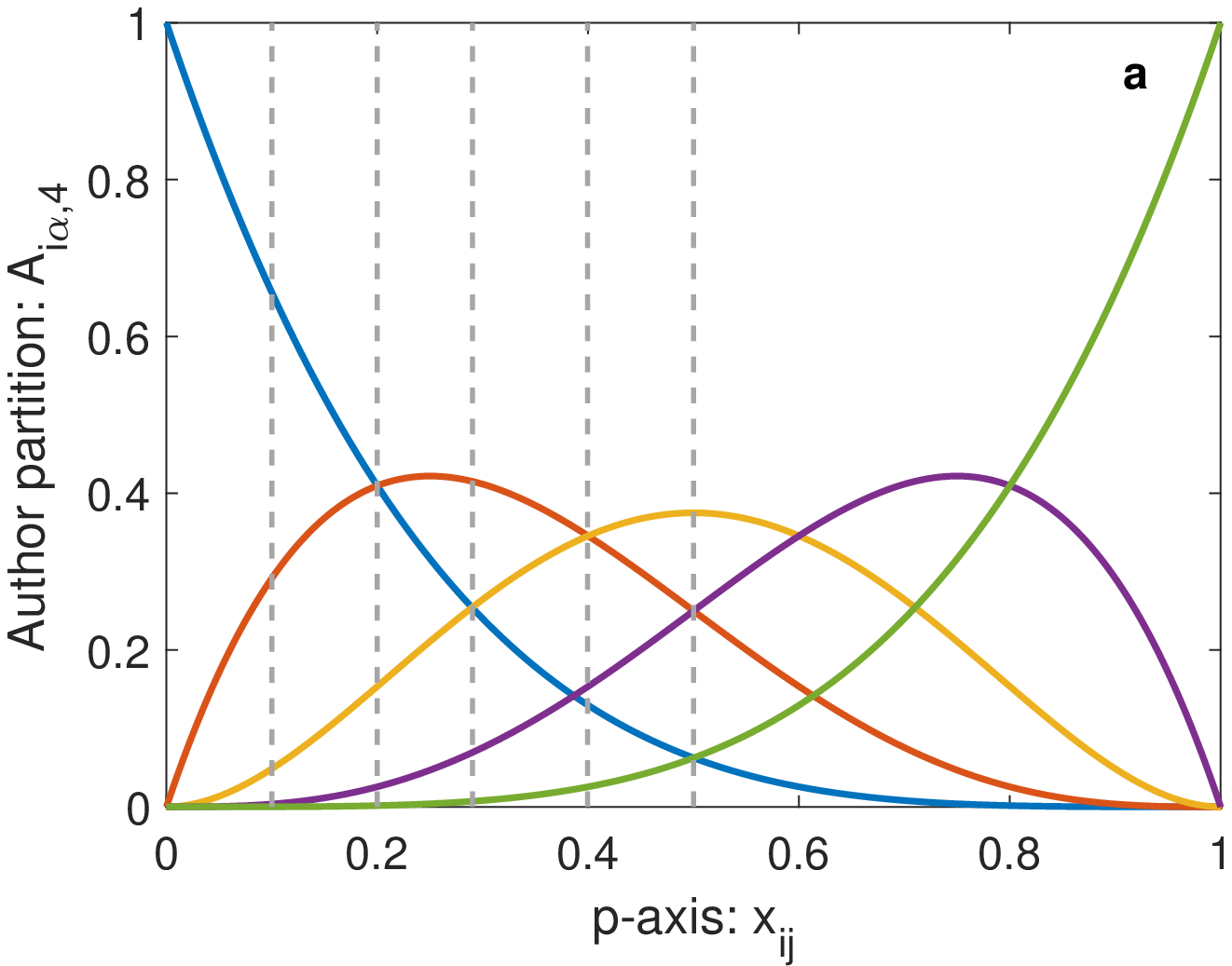}
 \hspace{-5mm}
 \includegraphics[width=9cm]{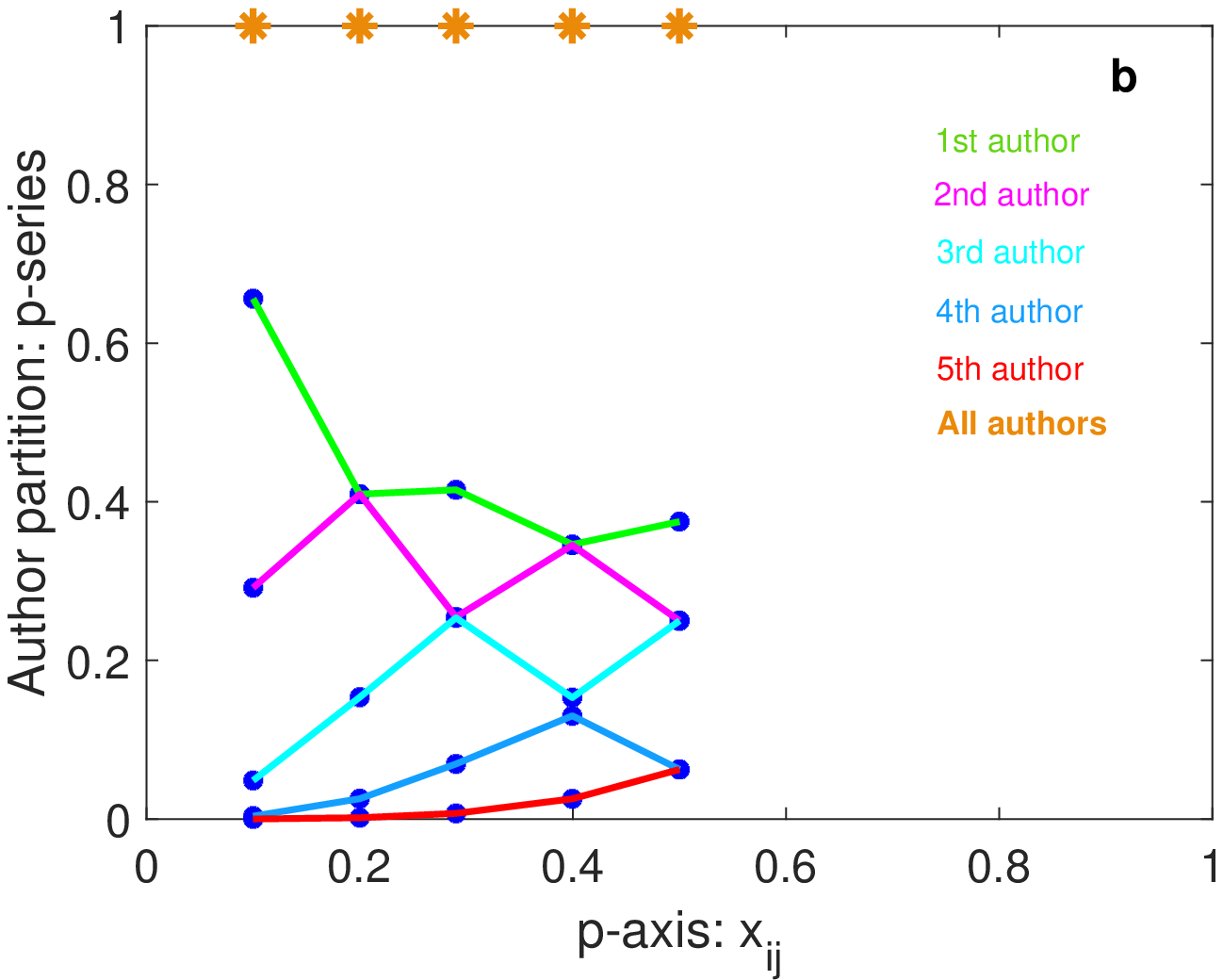}\vspace{-3mm}
  \end{center}
  \caption{Bernstein polynomials of degree four and different p-axes indicated by vertical dashed lines in gray (a), the p-sequences with different partitioning of author contributions (b). As the p-axis changes, note that partitioning amongst the authors changes. Some authors gain, while others lose. It reveals fascinating, complex dynamics of the author partitionings with strong re-shuffles.}
  \label{Fig_1}
\end{figure}
\begin{figure}[htp!]
\begin{center}
\vspace{-4mm}
\hspace{5mm}
 \includegraphics[width=8.69cm]{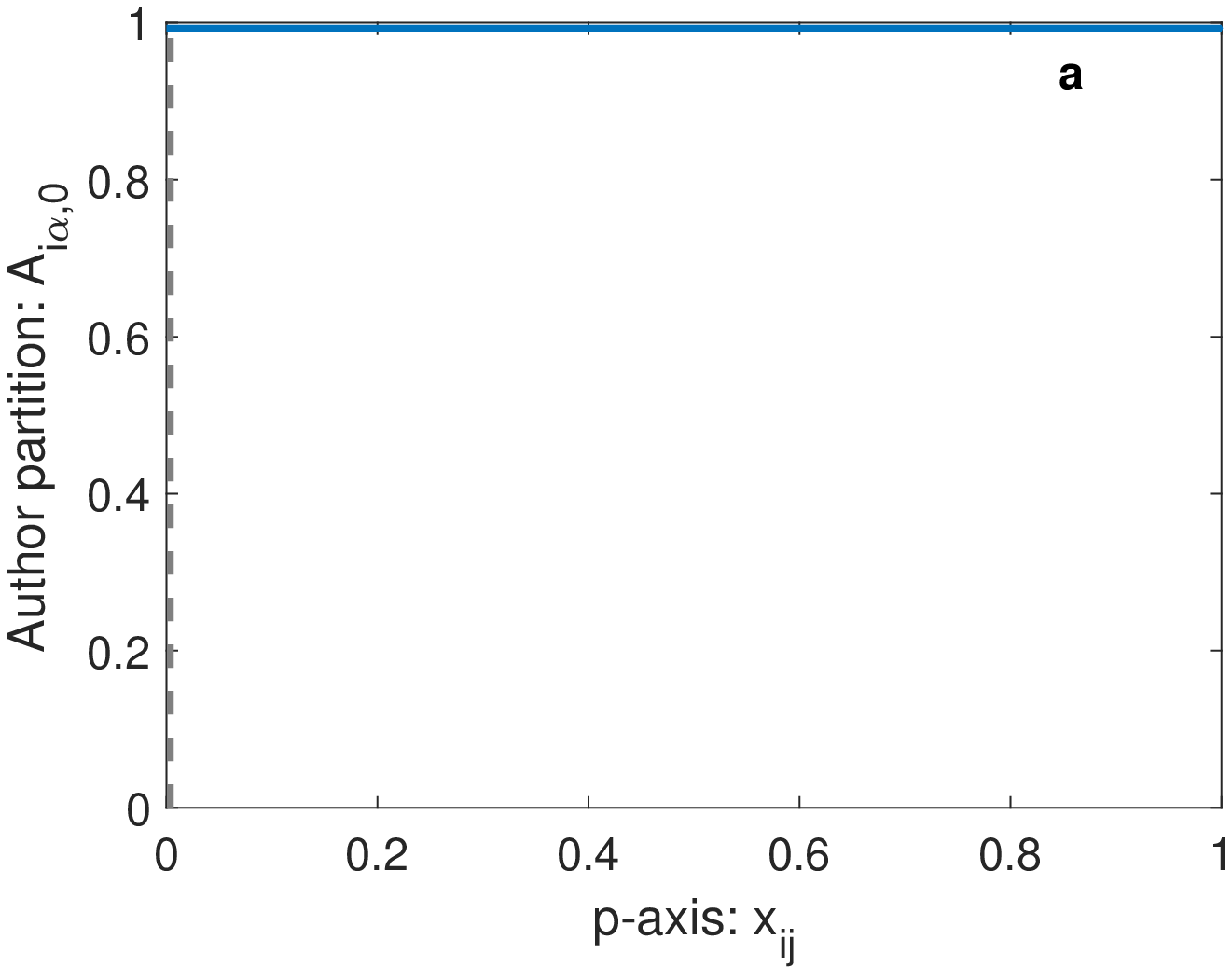}\hspace{-3mm}
 \includegraphics[width=8.69cm]{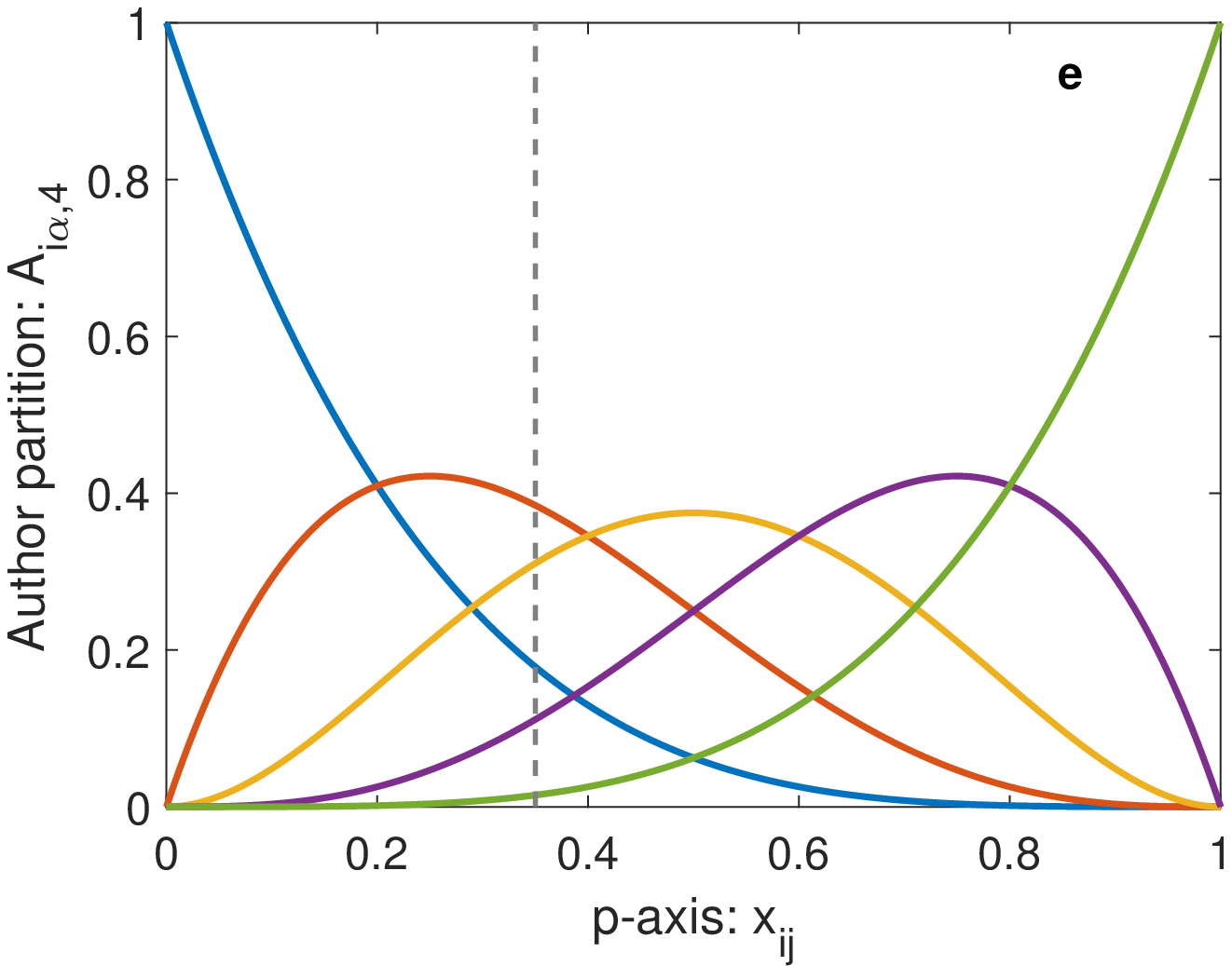}\\[-6.50mm]
 \hspace{5mm}
 \includegraphics[width=8.69cm]{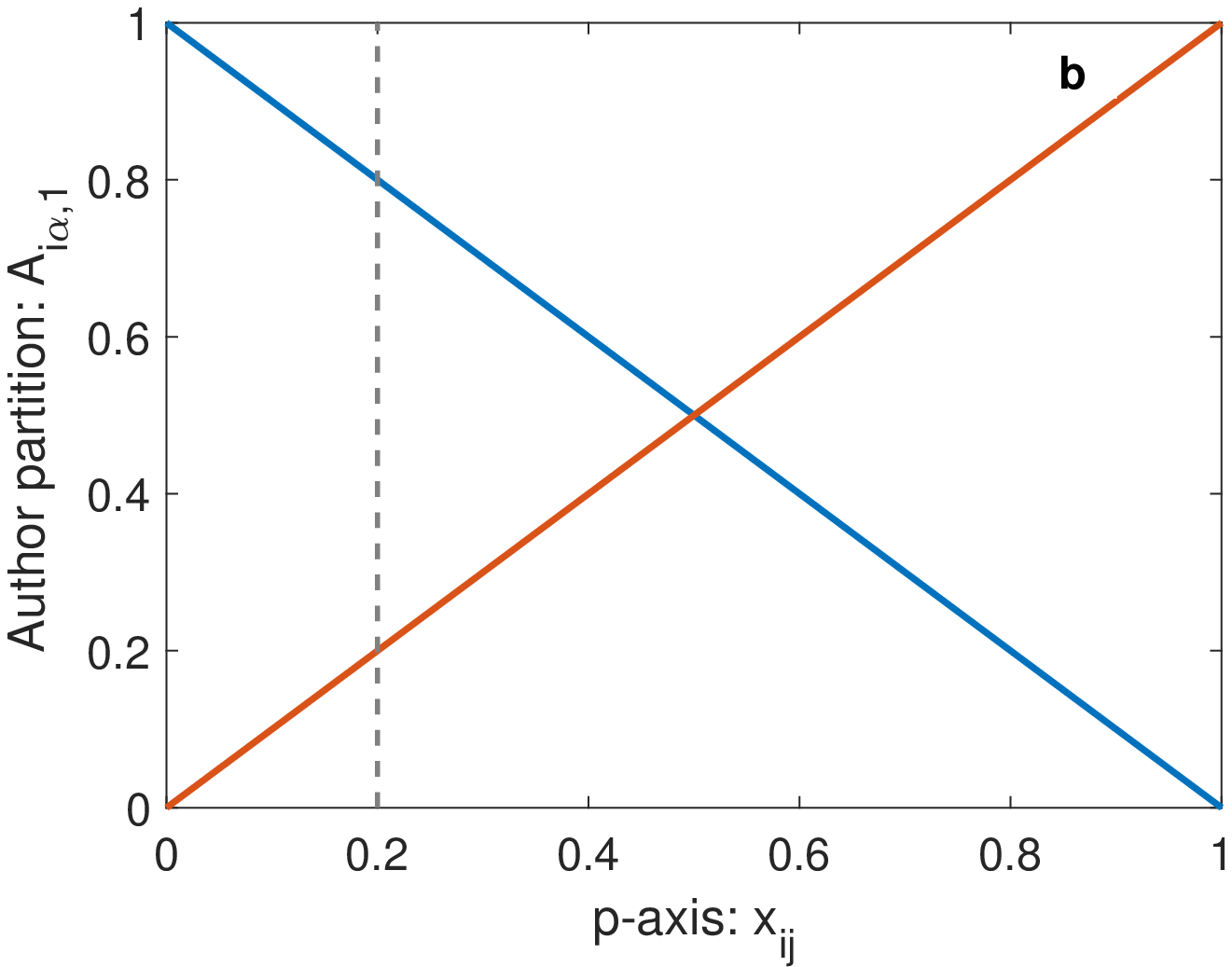}\hspace{-3mm}
 \includegraphics[width=8.69cm]{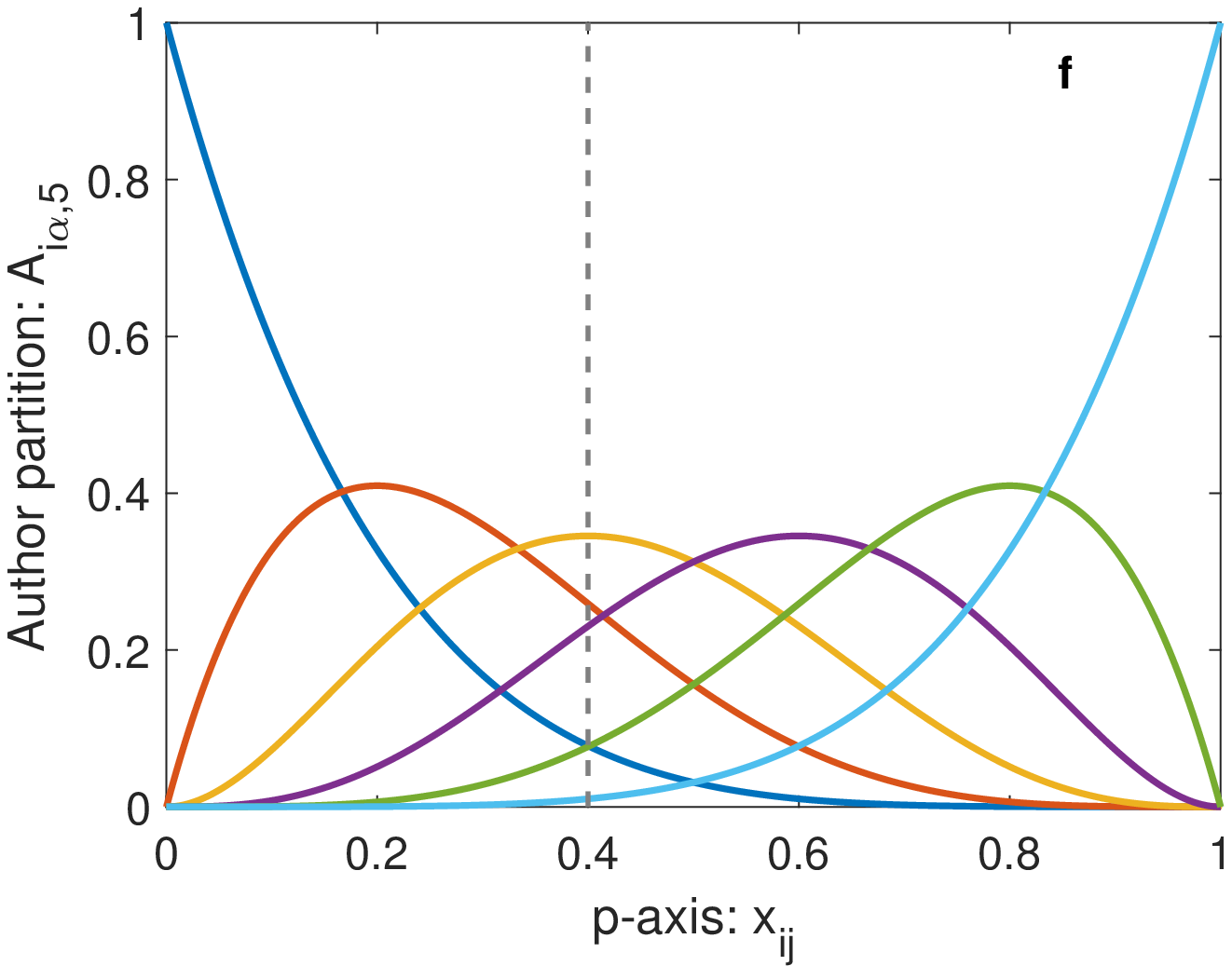}\\[-6.50mm]
 \hspace{5mm}
 \includegraphics[width=8.69cm]{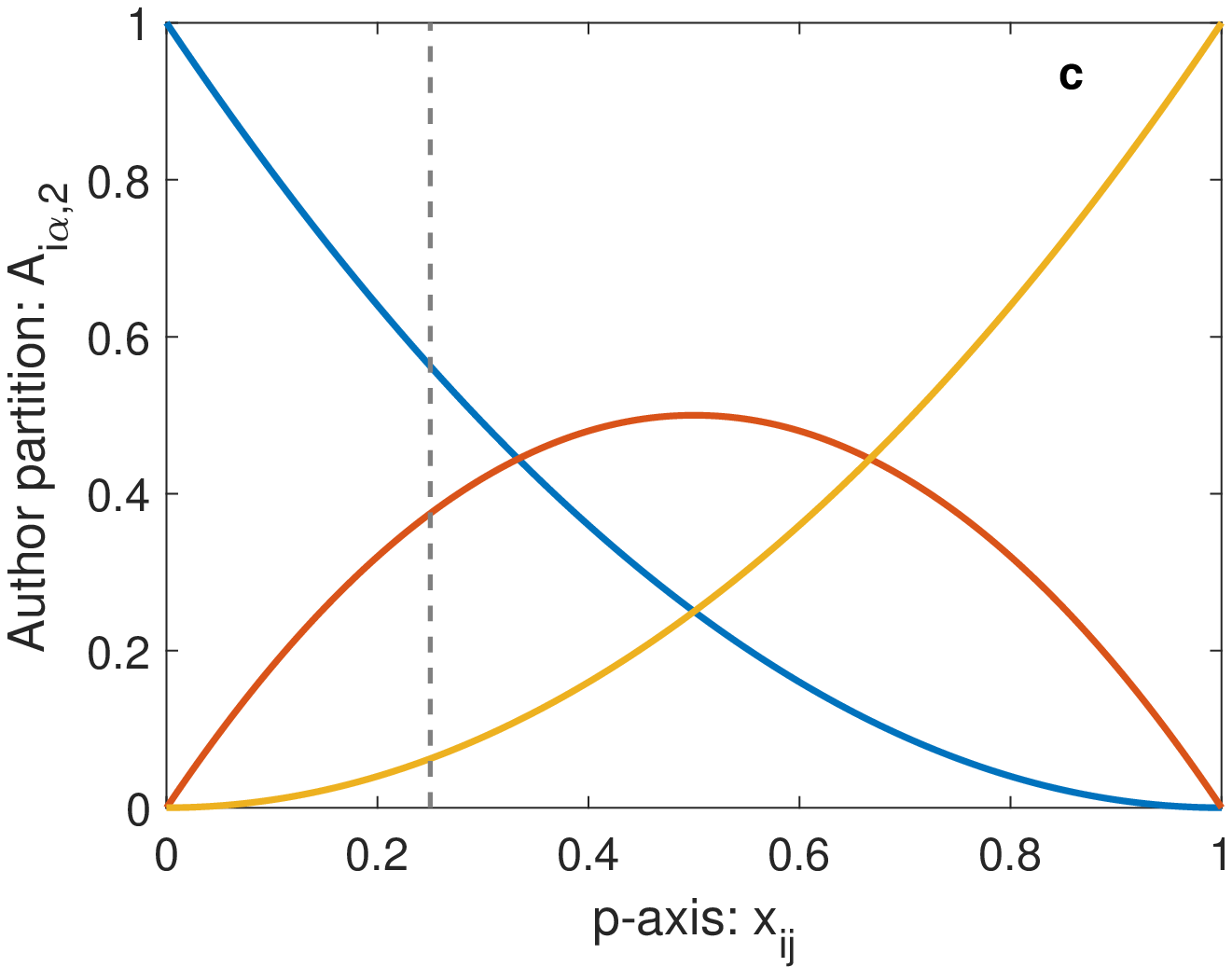}\hspace{-3mm}
 \includegraphics[width=8.69cm]{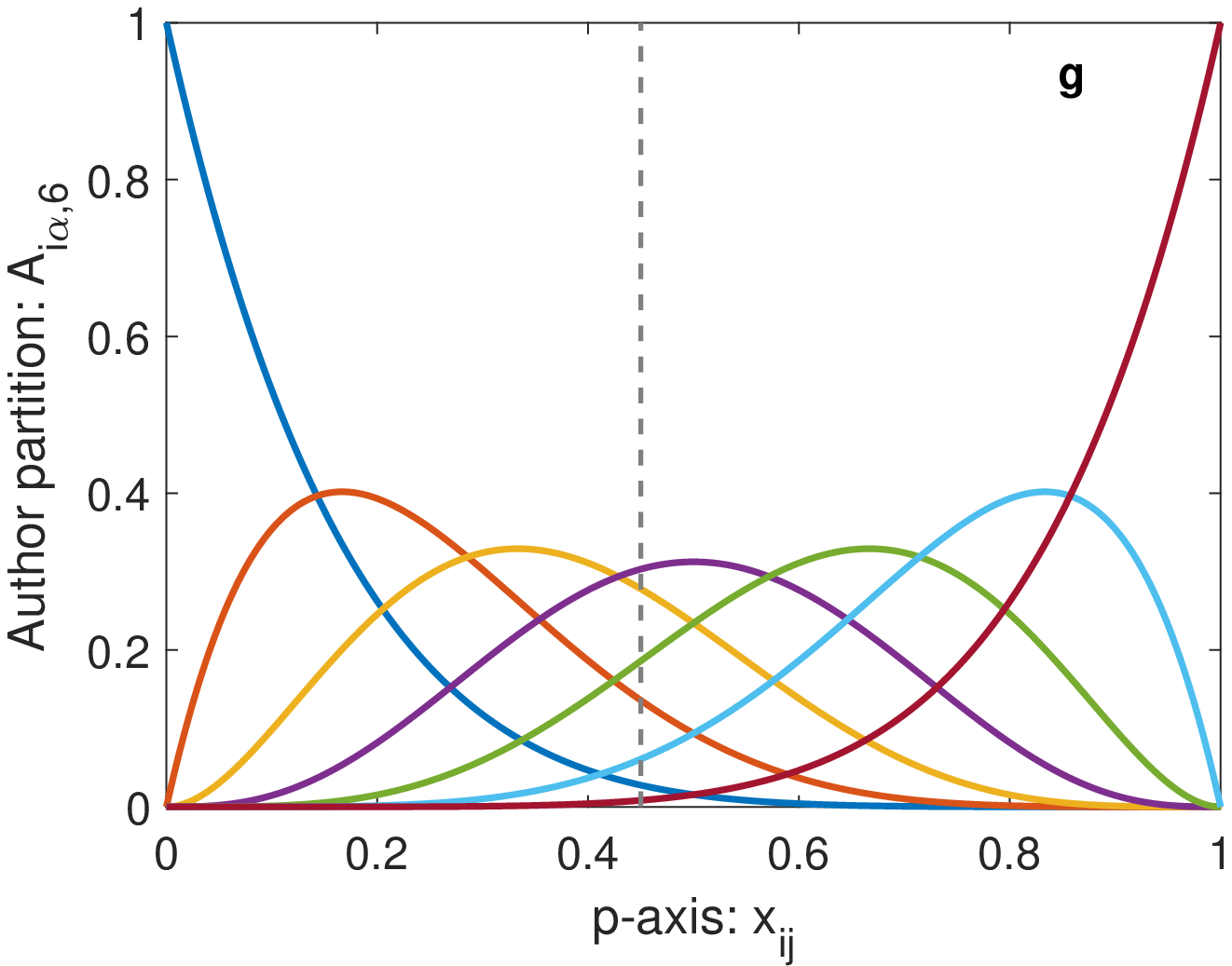}\\[-6.50mm]
 \hspace{5mm}
 \includegraphics[width=8.69cm]{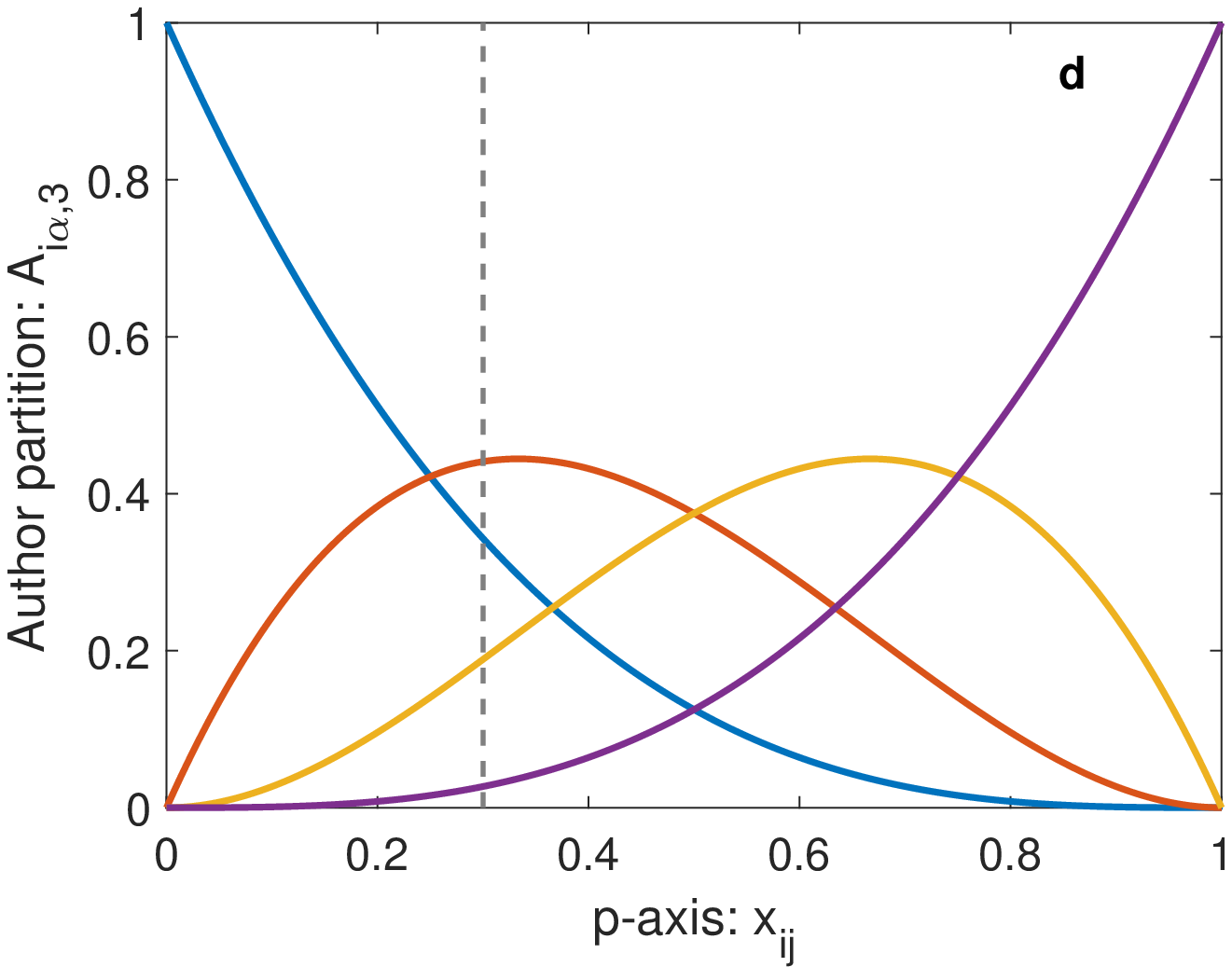}\hspace{-3mm}
 \includegraphics[width=8.69cm]{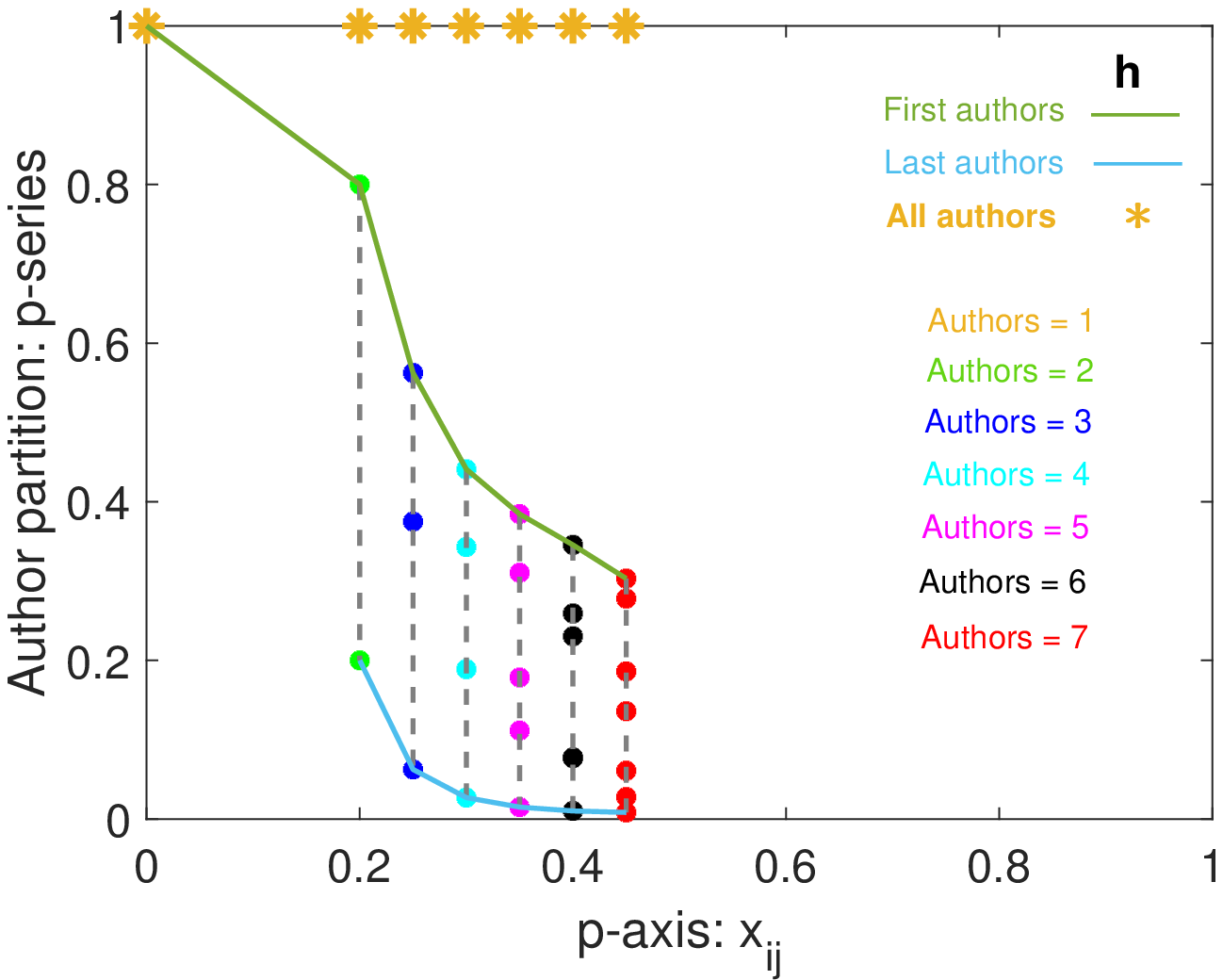}\vspace{-5mm}
  \end{center}
  \caption{Bernstein polynomials of degrees zero to six, partitioning the author contributions and the p-axes (a-g), and the p-sequences of author contribution partitions (h). By adding further co-authors, both the first and the last authors lose their earnings, however, the leading author loses the most, and most rapidly.}
  \label{Fig_2}
\end{figure}
\begin{multicols}{2}
Importantly, depending on the choice of $p^a_{i_j} \!\!=\! x^a_{i_j}$ (which means shifting the p-axis to the right or to the left), we can have quite different partitions of unity with $\mathcal A_{i_{\alpha, j}}$. The p-axis closer to the origin refers to the very high contribution of the first author and substantial to considerable contributions of the few following co-authors. Yet, the contributions of remaining co-authors can be minimal or (almost) negligible. If we move to the right, the contribution of the first author decreases rapidly, and that decreased value is distributed proportionally to all other co-authors, particularly the first following co-authors gain substantially, but, the rest of the co-authors do not gain that much, or only get negligible benefits. As the p-axis moves closer to the center, i.e., $p^a_{i_j} \approx 0.5$, it manifests more uniform and closer contributions among the co-authors. Some authors can even have equal contributions. This also includes the shared first authorship, or shared other intermediate authorships in the byline. By moving the p-axis to the right, some authors gain (lose) while in reciprocity, others lose (gain). Below, I present two examples to visualize these dynamical aspects.
\\[3mm]
For a situation with $j = 4$ (total 5 authors), the results are presented in Fig. \ref{Fig_1}. The main idea here is to show how for the same number of authors there can be fundamentally different author contribution partitions.
Figure \ref{Fig_1}a displays the five Bernstein polynomials and the five differently positioned p-axes from left to right 
$[p^1_{i_{4}}, p^2_{i_{4}}, p^3_{i_{4}},  p^4_{i_{4}}, p^5_{i_{4}}]
= [0.1, 0.2, 0.2903, 0.3994, 0.5]$ ($a = 1, 2, 3, 4, 5$ indicate different locations of the p-axis).
Each of these p-axis intersects the polynomials at five distinct points. When the resulting ordinate values are collected and arranged in decreasing orders, we obtain author partitions for five authors:\\[3mm] 
$\mathcal A_{i_{\alpha,4}} \left ( x^1_{i_4}\right ) = [0.6561, 0.2916, 0.0486, 0.0036, 0.0001]$,\\[2mm]
$\mathcal A_{i_{\alpha,4}} \left ( x^2_{i_4}\right ) = [0.4096, 0.4096, 0.1536, 0.0256, 0.0016]$,\\[2mm]
$\mathcal A_{i_{\alpha,4}} \left ( x^3_{i_4}\right ) = [0.4151, 0.2547, 0.2537, 0.0695, 0.0071 ]$,\\[2mm]
$\mathcal A_{i_{\alpha,4}} \left ( x^4_{i_4}\right ) = [0.3461, 0.3453, 0.1531, 0.1301, 0.0254 ]$,\\[2mm]
$\mathcal A_{i_{\alpha,4}} \left ( x^5_{i_4}\right ) = [0.3750, 0.2500, 0.2500, 0.0625, 0.0625 ]$.\\[3mm]  
I call these the p-sequences of author partitions.
Interestingly, for the same number of authors, there can be quite different author contribution partitions. The fascinating dynamics is observed in Fig. \ref{Fig_1}b. The position of the p-axis completely re-shuffles the author contribution partition. By shifting the p-axis to the right, in general, the earnings of the first author decrease substantially. The intermediate authors earnings can be increased or decreased, but (here) the last author gains steadily, though that person's contribution remains relatively small (about 6\%). As the p-axis shifts to $x^2_{i_4} = 0.2$, the first and second authors emerge as shared first-author, and obtain the same earning, which is more than the others. As we continue to push the p-axis to the right, other intermediate authors may acquire common contribution status. At $x^3_{i_4} = 0.2903$, due to the new shared authorship, the second and the third authors gain the same earning, which is lower than the first author but higher than the following authors. An even more interesting situation appears at $x^4_{i_4} = 0.3994$ where again the first and the second authors come together to regain their shared first authorship and get a common earning. In addition, the third and the fourth authors obtain the second shared authorship status and garner equal earning, lower than the first shared author, yet, higher than the last author. Moreover, as the p-axis reaches the central location, $x^5_{i_4} = 0.5$, the second and the third, and the fourth and the fifth authors, respectively, achieve the same author contribution values as pairs of shared authors. So, the p-axis plays a dominant role in constituting the author contribution partitioning.
\\[3mm]
Now, I discuss the Bernstein polynomials and the corresponding partitions of unity for articles with 1 up to 7 authors. The idea here is to manifest how the author contribution partitions change drastically as the number of authors increases, and what it means for the control of the author number to an article such that the authors with genuine contributions properly get their earnings. 
 Depending on the nature of contributions among the authors (also the disciplines and/or institutions), we must suitably select $x^a_{i_j}$, as it plays the crucial role. Here, for demonstrative purposes, without loss of generality, I suggest a criterion for a specific discipline for up to seven authors. For this, select $x^a_{i_j}$ as follows: 
$[x^1_{i_0}, x^2_{i_1}, x^3_{i_2}, x^4_{i_3}, x^5_{i_4}, x^6_{i_5}, x^7_{i_6}] = [0.00, 0.20, 0.25, 0.30, 0.35, 0.40, 0.45]$. The corresponding partitions of unity and the p-sequences for the single author, 2 authors, 3 authors, ..., and 7 authors, respectively, are:
\\[3mm]
{\small
$\mathcal A_{i_{\alpha, 0}}\left ( x^1_{i_0}\right ) = [1.]$,\\[2.5mm]
$\mathcal A_{i_{\alpha, 1}}\left ( x^2_{i_1}\right ) = [.8, .2]$,\\[2.5mm]
$\mathcal A_{i_{\alpha, 2}}\left ( x^3_{i_2}\right ) = [.5625, .3750, .0625]$,\\[2.5mm]
$\mathcal A_{i_{\alpha, 3}}\left ( x^4_{i_3}\right ) = [.441, .343, .189, .027]$,\\[2.5mm]
$\mathcal A_{i_{\alpha, 4}}\left ( x^5_{i_4}\right ) = [.3845, .3105, .1785, .1115, .0150]$,\\[2.5mm]
$\mathcal A_{i_{\alpha, 5}}\left ( x^6_{i_5}\right ) = [.3456, .2592, .2304, 0.0777, .0768, .0102]$,\\[2.5mm]
$\mathcal A_{i_{\alpha, 6}}\left ( x^7_{i_6}\right ) = [.3032, .2779, .1861, .1359, .0609, .0277, .0083]$.
}
\\[3mm]
The 1 to 7 polynomials and p-axes are shown in Fig. \ref{Fig_2}a-g. Similarly, the 7 p-sequences of author partitions are displayed in Fig. \ref{Fig_2}h. We observe that, as the number of authors increase, their contributions get closer, something which is probably reasonable. In each of $\mathcal A_{i_{\alpha, j}}$, there is a unique value associated with the particular author $A$, depending on $A$'s position in the byline in that multiauthored article. 
 The P-Index for $A$ can be computed by combining the values for $\mathcal A_{i_{\alpha, j}}$ with equations (\ref{Eqn_2PQ}) and (\ref{Eqn_2P}).
Later, I implement this rule using real citation data from the Web of Science for a research scientist in order to infer a plausible \mbox{P-Index}.
\\[3mm]
Figure \ref{Fig_2}h reveals many astonishing features of the author partitioning as determined by the p-axis.  
By adding further co-authors, both the first and the last authors lose their earnings. However, the leading author loses the most, and most rapidly. As the number of authors increases, the author contributions are more close, but their relative differences change quite dynamically, some come closer to the neighboring authors than the others. For some p-axes, the author partitioning is such that the two (or more) neighboring authors get very close and practically obtain the same values. Multiple localized pairs (or bundles) of shared authors of different levels may appear with virtually the same partitioning values. This can be very useful when two or more authors have similar contributions such that we can formally assign them the appropriate citation values. 
\\[3mm]
The results presented in Fig. \ref{Fig_1}b and Fig. \ref{Fig_2}h show that depending on the number of authors and the partitioning scheme (the p-axis), sometimes the author contributions are more homogeneous, while in other situations, the author contributions are more heterogeneous. All these are plausible scenarios that we may encounter in reality. In general, the partitioning reveals a complex pattern and dynamics.
\\[3mm]
A particularly important aspect is represented by the envelope of the Bernstein polynomials (Mabry, 2003):
\begin{equation}
\mathcal E_j(x_{i_j}) = \frac{1}{\sqrt{2 \pi j ~x_{i_j}\left ( 1- x_{i_j}\right) }},
\label{BernsteinPolyEn}
\end{equation}
which, for all $j+1$ authors, runs over the local maxima of polynomials. Take a hyperauthorship paper $i$ with 1001 authors. If the first author has contributed largely (a situation when $x_{i_j}$ is closer to 0), then, from Fig. \ref{Fig_2} we can infer that, except for the first following authors, all remaining author's actual makings from this paper are negligible. In another scenario, if there are more closer contributions among the authors (a situation when $x_{i_j}$ is closer to 0.5), then the nearest point on $\mathcal E_j(x_{i_j})$ for first author is ${1}/{\sqrt{\pi 1000/2}} \approx 0.0252$. This is an upper bound for the first author, but very close to that individual's actual contribution. This limit decreases continuously as the number of authors increases. Assume that there are 5000 citations for the paper $i$. This means, even for the first author, the actual earning is about 126. For some following authors in the byline, their earnings reduce steadily. However, after that, the remaining authors in the middle and lower positions in the byline will have insignificant earnings. This shows that unnecessarily adding authors in an article is not of actual benefit to anyone, the authors, journals, and institutions. 
\\[3mm]
I am now in the position of presenting an automated author contribution partitioning algorithm, and I call this the p-algorithm.
Here is the clear five-step algorithm: 
I. Select partitioning functions (such as the Bernstein polynomials as adopted here, but one is free to suitably choose any). 
II. Fix the p-axis, $p^a_{i_j} = x^a_{i_j}$. 
III. Collect all the ordinate points 
$\Big\{ \mathcal A_{i_{\alpha, j}}\left( x^a_{i_j}\right), \alpha = 0, 1, 2, ...,j\Big\}$ 
that are produced as the p-axis intercepts the partitioning functions. 
IV. Arrange these points in descending orders 
$\big\{\mathcal A_{i_{\alpha, j}}\!: \mathcal A_{i_{\alpha, j}} =\mbox{descend}\left(\mathcal A_{i_{\alpha, j}}\right)\big\}$ to obtain the partition of author contributions. 
V. Apply $\{\mathcal A_{i_{\alpha, j}}\}$ to (\ref{Eqn_2PQ}) and obtain the P-Index from (\ref{Eqn_2P}).
\\[5mm]
{\bf \large Essence of the P-Index}
\\[0mm]
The P-Index is the personal citation metric. It is personal to $A$ because, it only gathers $A$'s real earnings based on the author's contributions to constitute the author personal metric. The P in (\ref{Eqn_2P}), in fact, is the true measure of $A$'s earning from all the articles in which $A$ is an author, and the outcomes from those articles. The main essence lies in the author partitioning $\mathcal A_{i_{\alpha, j}}$. An author, publishing more impactful articles, containing less co-authors and contributing with high numerically measurable values (e.g., assigned with respect to the very first positions in the byline) obtains a high P-Index. This also motivates researchers not to include unnecessary co-authors in an article, and publish neat and potentially impactful articles. 
Moreover, someone with negligibly minimal contribution may even prefer not to be a co-author in the article $i$, because, this potentially adversely impacts that person's P-Index.
These have huge implications in the P-Index, but also to the relevant journals as well as to the institutions. 
\\[3mm]
The P-Index is based on one's productivity (number of articles, $N$), personal contributions to multiauthor articles (contractions, ${\mathcal A_{i_{\alpha, j}}}$), and the impact (citations, $C_i$). So, it is fundamentally different from the H-Index, {and any of its variants and improvements (Galam, 2011)}. The three important parameters, the productivity, contribution, and the impact are clearly defined and fully employed in the P-Index. However, the H-Index sorts out $C_i$ with decreasing values and finds the largest number $H$ in $i$ such that $C_i \ge H$. It does not use the entire productivity, and personal contributions are not discerned at all. 
\\[3mm]
{The P-Index can be perceived as a generalization of the journal impact factor (IF, Web of Science, 2022) for a journal $\mathcal J$ over all the articles $i$ published in it in the last two (or five) years, but without the contraction, i.e., setting $\mathcal A_{i_{\alpha, j}} =1$, because an article is associated with only one journal. However, for the person $A$, the contraction $\mathcal A_{i_{\alpha, j}}$ plays an eminent role. As the P-Index is associated with the person $A$, we can also call it the global dynamic Personal-ImpactFactor, or simply P-IF. Global, because, it is for the entire duration since the publication of the first paper by $A$ until present. Yet, we can apply a local time window to (\ref{Eqn_2P}) to obtain the Personal-ImpactFactor in that time duration. This implies that, essentially, we can view the P-Index as a generalization of the CiteScore (Scopus, 2022). Additionally, the globalness also refers to the fact that the Personal-ImpactFactor is influenced by $A$'s all publications and citations, whereas the H-Index is concerned only with the relatively highly cited papers. Moreover, contrary to the H-Index, which is non-decreasing, the Personal-ImpactFactor is dynamic, because it evolves over time, and depending on $A$'s research activities and impact, it can increase or decrease. This explains the representativeness, compactness and dynamics of the P-Index. This may also imply that, for the sake of everyone's benefit, probably no one wishes to include someone in the byline with negligible contribution.} 
\\[5mm]
{\bf \large Enhancement}
\\[0mm]
Consider the functions:
\begin{equation}
\mathcal A_{i_{\alpha, j}} \left ( x_{i_j}, s\right)= 
\frac{1}{s^j}
\begin{pmatrix}
j\\
\alpha
\end{pmatrix}
x_{i_j}^{\alpha} \left ( s - x_{i_j}\right )^{j-\alpha},
\label{Eqn_Partition_s}
\end{equation}
where $s > 0$. For $x_{i_j}\in [0, s]$, these polynomials form a partition of unity. I call them the Bernstein-S basis polynomials. When $s = 1$, the Bernstein-S basis polynomials become the Bernstein basis polynomials. Structurally, the Bernstein-S polynomials stretch (or contract, both the domain and polynomials) the Bernstein polynomials. Although the Bernstein polynomials can be recovered from these polynomials with the scaling $\chi_{i_j} = x_{i_j}/s$ there are technical advantages of the Bernstein-S polynomials over the Bernstein polynomials. The Bernstein-S polynomials offer us a much wider spectrum of author partitionings than the Bernstein polynomials. At a given (common) location $x$, the Bernstein-S polynomials provide a completely different partition of unity, and thus the p-series, that may be preferable than the p-series obtained from the Bernstein polynomials. By properly choosing the stretching parameter $s \neq 1$, we can substantially increase or decrease the author partition values of the first and the last authors, so are the partitions of the intermediate authors. The author partitions can get farther apart, or come closer, providing flexible and more legitimate values than those obtained from the Bernstein polynomials. However, the choice of the stretching parameter $s$ might differ from discipline to discipline, and should be decided by the authors, relevant institutions, or national or international entities. 
\\[3mm]
Here, I provide two examples for a paper with three authors with $s = 2$ (stretching), and $s = 0.55$ (contracting) domains and polynomials. For both, the partitionings are taken at the p-axis, $p^a_{i_j} = 0.25$ (vertical dashed line in gray). First, Fig. \ref{Fig_s}a displays both the Bernstein polynomials and the stretched Bernstein-S polynomials, their corresponding partitionings $[0.5625, 0.3750, 0.0625]$ (square symbols in cyan) and $[0.7656, 0.2188, 0.0156]$ (circle symbols in magenta), respectively. With the stretched {Bernstein-S} polynomials, the earning of the first author has increased largely probably reflecting that author's real contribution, while the second author loses substantially, and the third author loses significantly. Second, Fig. \ref{Fig_s}b shows both the Bernstein polynomials and the contracted Bernstein-S polynomials, their corresponding partitionings $[0.5625, 0.3750, 0.0625]$ (square symbols in cyan) and $[0.4959, 0.2975, 0.2066]$ (circle symbols in magenta), respectively. With the contracted Bernstein-S polynomials, the earning of the first author has decreased significantly, and the second author also loses substantially. However, the third author gains largely, likely mirroring that person’s real contribution. 
Contrary to the stretching, where the gap between the first and the second author became huge and the earnings among the authors depart away from each other making the distribution more heterogeneous, with contraction, the author contributions come closer, and become homogeneous. However, note that different p-axes provide completely different scenarios. This clearly demonstrates the practical usefulness and superiority of the Bernstein-S polynomials over the Bernstein polynomials.
\\[5mm]
{\bf \large A novel concept}
\\[0mm]
{As a quantitative modification of the H-Index, Galam (2011) presented the GH-Index, a denser author metric than the H-Index. The improved GH-Index is equivalent to the H-Index, but can differ drastically from the H-Index. It relies on Tailor Based Allocations (TBA) for multiple authorship. The allocations are tailored on the basis of each author contribution, and is the first method to do so. Several protocols to TBA are suggested. The proposed heterogeneous TBA favors the first and the last authors, strongly the last one. However, this method is discrete and contains two unknown parameters, making it an intricate process. Moreover, except for those favored authors, the contribution distribution among the other intermediate authors is homogeneous, and does not change substantially even for different choices of the additional parameters. Yet, the GH-Index cannot accommodate the shared authors. The P-Index is a completely novel, compact and unified concept that also rectifies existing shortcomings. Furthermore, the P-Index utilizes a simple, single system of smooth functions, and presents an automated, flexible and realistic method for partitioning author contributions.}
\\[5mm]
{\bf \large Implications}
\\[0mm]
In measuring one's total number of citations, the existing author metrices or measures (e.g., Web of Science, Scopus) simply set $\mathcal A_{i_{\alpha, j}}= 1$, neglecting the author contraction. Then, (\ref{Eqn_2}) becomes
\begin{equation}
C^{false} = C_{N_s}^s + \sum_{i = N_s+1}^{N} C_i.
\label{Eqn_4}
\end{equation}
However, this cannot represent the true measure of $A$'s total citations, and has thus been tagged as ``false". For an individual who has much fewer single author articles and has contributed only little to many multiauthored articles, $C \ll C^{false}$. This is a situation where the existing author indices can quite falsely measure one's actual contributions on citations {(Galam, 2011)}. Assume that a particular author $A_c$ is listed as coauthors of influential 25 articles and has the H-Index 25, but most of these articles have large number of co-authors, and $A_c$'s real contributions are very low, probably close to the end of the byline with a small numerical value in the partitioning. Although the H-Index names $A_c$ as a successful research scientist (scholar), the question is, does $A_c$ really deserve this? The answer is non-placet. The P-Index assigns $A_c$ the appropriate author metric, and removes the existing delusion. This will have huge implications in fair decision making by academic and scientific selection committees, award and prize nomination committees, funding grant recommendations (e.g., NSF-USA, ERC-EU, DFG-Germany, ANR-France, UKRI-UK), and institutional and personal recognitions. 
\\[3mm]
The author metric, or personal index, P-Index in (\ref{Eqn_2P}) strongly suggests that the literature should explicitly distribute the contributions in numerical values over all the co-authors, i.e., the partitioning with 
$\mathcal A_{i_{\alpha, j}}$. This is an urgent need that all academic and research institutions, concerned authorities, authors, editors and publishers should take an immediate step to fill this gap. Otherwise, no author metric will be complete and fully meaningful. This can be achieved in several ways with different norms as presented and discussed above. The academic and research institutions, and the relevant entities are strongly encouraged to follow (or develop their own) some clear numerical contributorship representations. However, the authors should be self inspired for this as they are primarily responsible for, and have the control over partitioning the contributions among themselves and its implementation. The author contribution partition, mentioned in (IV) of the p-algorithm above, should be clearly stated in the Author Contributions (or in Acknowledgment) section of the paper. Moreover, the journal editors and publishers can also monitor the implementation of the author contribution partitioning norms. It is expected that this is practical. Only this can provide a physically meaningful measure of the total citations $C$ for the particular author, and thus the author metric, the P-Index. 
\\[3mm]
Here, I demonstrate that the existing concept of number of articles and total citations cannot provide one's real measure of research strength. I give a real example on how the actual total citations $C$ can be drastically lower than the false total citations $C^{false}$, currently the only concept in use, and how representative, {compact} and realistic the P-Index is. I took the citation data of a research scientist $A$ from the Web of Science, as of June 10, 2022. The total citation for $A$ is 2394 and the H-Index is 28 with 66 articles, only 62 of which are cited. $A$ is the single author of 6 articles, and the first (lead) author of 25 articles. Moreover, the highest number of co-authored articles for $A$ are: 1 time 7 co-authors, 6 times 6 co-authors, and 5 times 5 co-authors. This information provides some ideas about the person’s state of collaborative research, and also the ability of being independent and lead researcher. Here, I use the p-algorithm with partitioning proposed above. With this, surprisingly, the actual (personal) total citations for $A$ is just $C = 1172$, which is much below the false total citations $C^{false} = 2394$. So, in usual situation, $C$ is expected to be exceptionally less than $C^{false}$. Even more important is the P-Index, which is merely $18$. 
This is much below the H-Index. It distinctly manifests that the P-Index is a realistic, comprehensive and representative author metric, even stronger and condensed than the H-Index. Moreover, unlike many other indices {(e.g., the GH-Index)}, the P-Index is not related, and complementary to the H-Index. It is clear, because the P-Index {is unified, compact and} only considers one's genuine contributions in the citation database {which are}
lacking in the existing author metrices. 
\\[5mm]
{\bf \large Conclusion}
\\[0mm]
I have provided an author-personal metric (the P-Index), a well-defined, well-justified, inclusive, compact and automated citation measure that is based on the total citations of an individual’s genuine research contributions and impact, which will benefit all authors, journals and institutions.
\\[-1mm]
\begin{itemize}
{\footnotesize
 \item[1.] Abambres, M., Arab, P., 2016. Citation indexes accounting for authorship order in coauthored research - review and new proposal. Sci. Technol. Libr. 35, 263-278. doi: 10.1080/0194262X.2016.1242450.
 
 \item[2.] Bernstein, S.N., 1912. Demonstration du theoreme de Weierstrass fondee sur le  calcul des probabilites. Comm. Soc. Math. Kharkov 13, 1-2.

\item[3.] Castelvecchi, D., 2015. Physics paper sets record with more than 5,000 authors. Nature.
https://doi.org/\\10.1038/nature.2015.17567.

\item[4.] Galam, S., 2011. Tailor based allocations for multiple authorship: a fractional gh-index. Scientometrics 89, 365-379. DOI 10.1007/s11192-011-0447-1.

\item[5.] Hirsch, J.E., 2005. An index to quantify an individual's scientific research output. Proceedings of the National Academy of Sciences, USA 102, 16569-16572.

\item[6.] ICMJE, 2018. Defining the role of authors and contributors: International Committee of Medical Journal Editors. http://www.icmje.org/recommendations/browse/roles\\[-0mm]
-and-responsibilities/defining-the-role-of-authors-and-\\contributors.html.

\item[7.] King, C., 2012. Multiauthor papers: onward and upward. ScienceWatch Newsletter. http://archive. science watch.com/newsletter/2012/201207/multiauthor\_papers/.

\item[8.] Lapidow, A., Scudder, P., 2019. Shared first authorship. Journal of the Medical Library Association 107, 618-620. https://doi.org/10.5195/jmla.2019.700.

\item[9.] Mabry, R., 2003. Problem 10990. Amer. Math. Monthly 110, 59.

\item[10.] McNutt, K., et al., 2018. Transparency in authors’ contributions and responsibilities to promote integrity in scientific publication. Proceedings of the National Academy of Sciences, USA 115, 2557-2560. doi: 10.1073/pnas.1715374115.

\item[11.] Scopus, 2022. https://www.scopus.com.

\item[12.] Tscharntke, T., Hochberg, M.E., Rand, T.A., Resh, V.H., Krauss, J., 2007. Author sequence and credit for contributions in multiauthored publications. PLoS Biol 5(1): e18. doi:10.1371/journal.pbio.0050018.

\item[13.] Verhagen, J.V., Wallace, K.J., Collins, S.C., Scott, T.R., 2003. QUAD system offers fair shares to all authors. Nature 426, 602.

\item[14.] Web of Science, 2022: https://www.webofscience.com/.
}
\end{itemize}

\end{multicols}

\end{document}